\documentclass[doublecol]{epl2} 

\newcommand{\bd}{\begin{document}}
\newcommand{\ed}{\end{document}}
\newcommand{\bc}{\begin{center}}
\newcommand{\ec}{\end{center}}
\newcommand{\bfr}{\begin{flushright}}
\newcommand{\efr}{\end{flushright}}
\newcommand{\lt}{\left}
\newcommand{\rt}{\right}
\newcommand{\vs}{\vspace}
\newcommand{\hs}{\hspace}
\newcommand{\beq}{\begin{equation}}
\newcommand{\eeq}{\end{equation}}
\newcommand{\lb}{\linebreak}
\newcommand{\pb}{\pagebreak}
\newcommand{\mb}{\makebox}
\newcommand{\fb}{\framebox}
\newcommand{\mc}{\multicolumn}
\newcommand{\ben}{\begin{enumerate}}
\newcommand{\een}{\end{enumerate}}
\newcommand{\bit}{\begin{itemize}}
\newcommand{\eit}{\end{itemize}}
\newcommand{\oln}{\overline}
\newcommand{\lefq}{\lefteqn}
\newcommand{\ba}{\begin{array}}
\newcommand{\ea}{\end{array}}
\newcommand{\beqa}{\begin{eqnarray}}
\newcommand{\eeqa}{\end{eqnarray}}
\newcommand{\beqas}{\begin{eqnarray*}}
\newcommand{\eeqas}{\end{eqnarray*}}
\newcommand{\bfg}{\begin{figure}}
\newcommand{\efg}{\end{figure}}
\newcommand{\bds}{\begin{displaymath}}
\newcommand{\eds}{\end{displaymath}}
\newcommand{\btb}{\begin{tabbing}}
\newcommand{\etb}{\end{tabbing}}
\newcommand{\para}{\parallel}
\newcommand{\pad}{\partial}
\newcommand{\nn}{\nonumber}
\newcommand{\la}{\leftarrow}
\newcommand{\ra}{\rightarrow}
\newcommand{\lgla}{\longleftarrow}
\newcommand{\lgra}{\longrightarrow}
\newcommand{\La}{\Leftarrow}\newcommand{\Ra}{\Rightarrow}
\newcommand{\Lra}{\Leftrightarrow}
\newcommand{\Lgla}{\Longleftarrow}
\newcommand{\Lgra}{\Longrightarrow}
\newcommand{\lan}{\langle}
\newcommand{\ran}{\rangle}
\renewcommand{\a}{\alpha}
\renewcommand{\b}{\beta}
\newcommand{\g}{\gamma}
\newcommand{\G}{\Gamma}
\renewcommand{\d}{\delta}
\newcommand{\eps}{\epsilon}
\newcommand{\Th}{\Theta}
\newcommand{\s}{\sigma}
\newcommand{\lam}{\lambda}
\newcommand{\D}{\Delta}
\newcommand{\ds}{\displaystyle}
\newcommand{\vare}{E}
\newcommand{\pr}{\prime}
\newcommand{\ro}{\rho}
\newcommand{\nab}{\nabla}
\newcommand{\m}{\mu}
\newcommand{\n}{\nu}
\newcommand{\Sg}{\Sigma}
\newcommand{\p}{\pi}
\newcommand{\R}{I\!\!R}
\newcommand{\om}{\omega}
\newcommand{\Om}{\Omega}
\newcommand{\ovra}{\overrightarrow}
\newcommand{\ze}{\zeta}
\newcommand{\vart}{\vartheta}
\newcommand{\tri}{\triangle}
\newcommand{\f}{\frac}
\newcommand{\iny}{\infty}
\newcommand{\pro}{\propto}
\renewcommand{\arraystretch}{1.25}
\usepackage{booktabs,amsmath}
\usepackage{bm,amsmath,amssymb,amsfonts,latexsym,psfrag}
\usepackage{amssymb}
 \usepackage{array}
\usepackage{mathtools}
\usepackage{amsthm}
\usepackage{newlfont}
\usepackage{esdiff}
\usepackage{multirow}
\usepackage{graphics}
\usepackage{graphicx}
\usepackage{color}
\usepackage{tabularx}
\usepackage{bm}
\usepackage{url}
 \usepackage[colorlinks=true,citecolor=cyan,urlcolor=blue,bookmarks=true,bookmarks=true,bookmarksopen=true,bookmarksnumbered=true,bookmarksopenlevel=3]{hyperref}

\definecolor{airforceblue}{rgb}{0.36, 0.54, 0.66}
\definecolor{steelblue}{rgb}{0.27, 0.51, 0.71}
\definecolor{amber}{rgb}{1.0, 0.49, 0.0}

\title{Non-Commutativity effects in the  Dirac equation in crossed electric and magnetic fields.}
\shorttitle{Non-Commutativity effects in the  Dirac equation ...} 

\author{D.~Nath\inst{1}\and M. Presilla\inst{2,3}\and O.~Panella\inst{4} \and P. Roy\inst{5}}
\shortauthor{D. Nath  \etal}

\institute{                    
  \inst{1}Department of Mathematics, Vivekananda College, Kolkata-700063, India\\
  \inst{2}Dipartimento di Fisica e Astronomia ``G. Galilei''  Universit\`{a} di Padova, Padova, Italy\\
  \inst{3}Istituto Nazionale di Fisica Nucleare, Sezione di Padova, Padova, Italy\\
  \inst{4}Istituto Nazionale di Fisica Nucleare, Sezione di Perugia, Via A. Pascoli, Perugia, Italy\\
  \inst{4}Physics and Applied Mathematics Unit, Indian Statistical Institute, Kolkata-700108, India
}

\abstract{
In this paper we present exact solutions of the Dirac equation on the non-commutative plane in the presence of  crossed electric and magnetic fields. In the standard commutative plane such a system is known to exhibit contraction of Landau levels when the electric field approaches a critical value.  In the present case we find exact solutions in terms of the non-commutative parameters $\eta$ (momentum non-commutativity) and $\theta$ (coordinate non-commutativity) and provide an explicit expression for  the Landau levels. We show that non-commutativity preserves the collapse of the spectrum. We provide a dual description of the system: (i) one in which at a given electric field the magnetic field is varied and the other (ii) in which at a given magnetic field the electric field is varied. In the former case we find that momentum non-commutativity  ($\eta$) splits the critical magnetic field into two critical fields while coordinates non-commutativity ($\theta$) gives rise to two additional critical points not at all present in the commutative scenario.}

\begin{document}

\maketitle

\section{Introduction}
\label{Intro}
In recent years the $(2+1)$ dimensional Dirac equation has attracted a lot of attention because it is used to describe the motion of charge carriers in graphene and other Dirac materials \cite{novo}. In particular, the massless $(2+1)$ dimensional Dirac equation in the presence of a homogeneous magnetic field has many applications in graphene, for example creating bound states \cite{bound1,bound2}, studying quantum Hall effect \cite{a1}, fractional Hall effect \cite{a5}, Berry phase \cite{a1,b1} etc. An interesting phenomena takes place when in addition to the magnetic field an uniform electric field (which is orthogonal to the magnetic field) is introduced. It was shown that the spacing between different Landau levels decreases when the electric field strength approaches a critical value \cite{Lukose} and finally at the critical value the different Landau levels coalesce to form a band. Such  a phenomenon takes place in other materials as well, for example in Weyl semimetals \cite{arjona}. Later the $(2+1)$ dimensional Dirac equation in crossed fields was studied by a number of authors for different reasons, for example, analyzing this system from the point of view of coherent states \cite{peres}, studying Landau levels \cite{ali1}, oscillation of magnetisation \cite{ali2}, Weiss oscillation \cite{ma}, de Haas-van Alphen effect \cite{de haas,de haas1} etc.
 
On the other hand there are indications from studies in string theory that space-time could be noncommutative \cite{nc1,nc2}. Subsequently a number of models e.g, the hydrogen atom \cite{h1,h2}, central field problem \cite{gamboa}, Landau problem \cite{landau1,landau2,landau3}, Aharonov-Bohm system \cite{ab} etc. were studied in order to determine the effect of non commuting coordinates and momenta on observables like the energy. In particular, various aspects of the Dirac equation in non commutative plane were examined in a number of papers \cite{ncd1,ncd2,ncd3,ncd4,ncd5,ncd6,ncd7,ncd8}. Among the different relativistic systems on the non commutative plane, graphene has been studied by a number of authors. For example, the effect of noncommutativity on Shubinkov-de Haas oscillations \cite{jellal}, Hall effect \cite{horvathy,jellal1,falomir,ncd3,santos} has been examined. It has also been shown that turbostratic graphene layers can be described as noncommutative $(2+1)$ space-time \cite{petit}. In this paper our objective is to further explore the effect of non commutativity on certain phenomena in graphene and similar materials. More specifically we shall study the noncommutative analogue of the model considered in ref\cite{Lukose}. In particular, we shall analyze the effect of non commutative coordinates and momenta on the spectrum.  To be more specific, we shall examine the system by (1) varying the magnetic field against a fixed electric field (2) varying the electric field against a fixed magnetic field. It will be seen that apart from critical points which have commutative counterparts, there are some critical points which exist only in the noncommutative scenario. 

\section{ {$(2+1)$} 
dimensional Dirac equation in crossed fields}
\label{21Dirac}
It is convenient to start with the manifestly covariant and time dependent Dirac equation in an external electromagnetic field obtained according to the minimal coupling requirement:
\begin{equation}
\label{21Diraceq}
\left[ \gamma^\mu \left( i\partial_\mu -\frac{q}{c} A_\mu\right ) -\Delta c\right]\Psi(x) =0\,.
\end{equation}
In 2 space dimensions we have $x^0=ct$, $x^1=x$, $x^2=y$, $\gamma^0=-i\sigma^x$, $\gamma^0=i\sigma^y$ and $\partial_\mu= \partial/\partial x^\mu$.  $A^\mu=(\phi, \bm{A})$ is the four-potential,  $A^0=\phi$ the scalar potential, $A^1= A_x$, $A^2=A_y$ for the vector potential $\bm{A}=(A_x,A_y)$ and $\Psi(x)$ is a two component spinor. 

Eq.~(\ref{21Diraceq}) can be straightforwardly put in the standard time-dependent form:
\begin{equation}
\label{timeindepDirac}
i\hbar \,\partial_0 \Psi = \left[ \bm{\sigma}\cdot \left(\bm{p}-\frac{q}{c}\bm{A}\right) + \frac{q}{c} A^0 +\sigma_z \Delta c\right]\Psi.
\end{equation}

 Assuming $\Psi= e^{-i\frac{\eps}{\hbar}t}\psi$ one derives the eigenvalue equation $H\psi=\eps\psi$:
\begin{equation}
\label{eigenvalue}\left[
v_F\bm{\sigma}\cdot \left(\bm{p}-\frac{q}{c}\bm{A}\right) + q \phi +\sigma_z \Delta v_F^2 \right]\,\psi= \eps\,\psi.
\end{equation}
In ref.~\cite{Lukose} the authors obtained the spectrum and wave-functions of  Eq.~(\ref{eigenvalue}) with a configuration of crossed external electric and magnetic fields. More precisely they considered a  uniform electric field along the +$x$ direction $\bm{E}=E\hat{\bm{i}}$ along with a magnetic potential in the (x,y) plane ${\bm{A}}=(-B/2\,{y},B/2\,{x})$. It was shown \cite{Lukose} that the electric field modifies the spectrum in such a way that the Landau levels collapse whenever the electric field exceeds a critical value. 

\subsection{Dirac equation in non commutative plane}
Here we would like to show that the system is also exactly solvable when considered within the realm of non commutative quantum mechanics. In the noncommutative plane, the coordinates and momenta satisfy the following commutation relations:
\begin{eqnarray}
\label{NCEQ}
\left[{\hat{x}},{\hat y}\right]&=&i\theta, \qquad\qquad[{\hat{p}}_x,{\hat{p}}_y]=i\eta,\nonumber \\ \left[{\hat{x}}_i,{\hat p}_j\right]&=&i\hbar\left(1+\frac{\theta\eta}{4\hbar^2}\right)\delta_{ij}, \qquad \theta,\eta \in \mathbb{R}\, .
\end{eqnarray}
The noncommutative Dirac equation equation corresponding to (\ref{eigenvalue}) reads \footnote{Keeping in mind possible applications of the results to graphene and other Dirac materials, we have replaced $c$ by the Fermi velocity $v_F$.}
\begin{equation}\label{d2}
\left[
v_F\bm{\sigma}\cdot \left(\bm{\hat p}-\frac{q}{c}{\hat{\bm{A}}}\right) + q {\hat\phi} +\sigma_z \Delta v_F^2 \right]\,\psi= \eps\,\psi.
\end{equation}
It is now necessary to specify the electromagnetic potentials. Choosing 
\begin{equation}
\hat{\bm{A}}=(-B/2\,\hat{y},B/2\,\hat{x})\, , \qquad  \hat{\phi}= -E\,\hat{x}\,,\qquad q=-e,
\end{equation}
we obatin from (\ref{d2})
\begin{eqnarray}\label{cdirac1}
&&\left[
v_F{\sigma}_x \left({\hat p}_x-\frac{e}{2c}B \hat{y}\right)+v_F \sigma_y \left( p_y+\frac{e}{2c}B \hat{x}\right)\phantom{xxxxxxx} \right.\nonumber \\
&&\phantom{xxxxxxxxxxxxx.}\left.\phantom{\frac{1}{2}}+ eE\hat{x}  +\sigma_z \Delta v_F^2 \right]\,\psi= \eps\,\psi.
\end{eqnarray}

In order to solve the above equation it is now necessary to express it in terms of the commuting variables. Using the Seiberg-Witten map the noncommuting variables $({\hat x},{\hat y},{\hat p}_x,{\hat p}_y)$ can be expressed in terms of the commuting coordinates and momenta as
\begin{equation}\label{c-nc}
\ba{l}
\ds{\hat x}=x-{\tilde\theta}p_y,~~~~{\hat p}_x=p_x+{\tilde\eta}y,\\\\
\ds{\hat y}=y+{\tilde\theta}p_x,~~~~{\hat p}_y=p_y-{\tilde\eta}x,~~~~{\tilde\theta}=\frac{\theta}{2\hbar},~~~~{\tilde\eta}=\frac{\eta}{2\hbar}.
\ea
\end{equation}
Now using (\ref{c-nc}) in (\ref{cdirac1}) we get
\begin{eqnarray}\label{psitilde}
&&\left[\ds \s_x\left((1-\tilde{B}\tilde{\theta})p_x-(\tilde{B}-\tilde{\eta})y\right)\right.\phantom{xxxxxxxxxxxxxxxx}\nonumber \\&&\phantom{xxxxxxx}\left.+\s_y\left((1-\tilde{B}\tilde{\theta})p_y+(\tilde{B}-\tilde{\eta})x\right) \right.\nonumber\\
&&\left. \ds\phantom{xxxxxxx}+\tilde{E}(x-\tilde{\theta}p_y)+\s_z \Delta v_F\right]\psi=\tilde{\eps}\psi,
\end{eqnarray}
where
\beq
\ds \tilde{B}=\f{eB}{2c},~~
\tilde{E}=\f{eE}{v_F},~~
\tilde{\eps}=\f{\eps}{v_F}.
\eeq  
  
\section{Solution}
Eq.~(\ref{psitilde}) is a coupled differential equation in terms of commuting coordinates and momenta. In a first step to solve this equation we now use the transformation
\beq
\psi(x,y)=e^{i\a xy}\phi(x,y),~~~~\a=\f{({\tilde B}-{\tilde\eta})}{(1-\tilde B\tilde\theta)^2}
\eeq
on (\ref{psitilde}) to obtain
\begin{eqnarray}
\label{psitilde2}
&&\left[ \s_x p_x +\s_y\left(p_y+B_0x\right)+\tilde{E}\g x -\f{\tilde{E}\tilde{\theta}}{1-\tilde{B}\tilde{\theta}}p_y \phantom{xxxxxxx}\right. \nonumber\\ && \left.\phantom{xxxxxxxxxxxxxx}+\s_z \f{\Delta v_F}{1-\tilde{B}\tilde{\theta}}\right]\phi
=\f{\tilde{\epsilon}}{1-\tilde{B}\tilde{\theta}}\phi,
\end{eqnarray}
where
\beq
B_0=\f{2(\tilde{B}-\tilde{\eta})}{1-\tilde{B}\tilde{\theta}},~~\\
\g=\f{1-2\tilde{B}\tilde{\theta}+\tilde{\eta}\tilde{\theta}}{(1-\tilde{B}\tilde{\theta})^2}.
\eeq
From (\ref{psitilde2}) it can be seen that the interactions are in the $x$ direction and the motion along the $y$ direction is free. Therefore we take the spinor $\psi(x,y)$ to be of the form
\beq\label{spinor1}
\phi(x,y)=e^{iky}\chi(x),
\eeq
where $k$ denotes the momentum in the $y$ direction. Now substituting (\ref{spinor1}) in (\ref{psitilde2}) we obtain
\beq \label{feq}
\left[\ds \s_x p_x +A\s_y+\Delta_0\s_z-(\tilde{\epsilon}_0-Fx)\right]\chi=0,
\eeq 
where
\beq
\ds\tilde{\epsilon}_0=\f{\tilde{\epsilon}+\tilde{E}\tilde{\theta}\hbar k}{1-\tilde{B}\tilde{\theta}},~~\\
A=\hbar k+B_0 x,~~\\
\Delta_0=\f{\Delta v_F}{1-\tilde{B}\tilde{\theta}},~~\\
F=\tilde{E}\g.
\eeq
It is easy to see that dependence of $A$ on $x$ makes it difficult to obtain the solution of (\ref{feq}). To circumvent this problem we take the spinor $\chi$ to be of the form
\beq\label{chi}
\chi=\left[\ds \s_x p_x +A\s_y+\Delta_0\s_z+(\tilde{\varepsilon}_0-Fx)\right]G(x).
\eeq  
Then from (\ref{feq}) and (\ref{chi}) we obtain
\beq
\left(\ba{cc} J+\f{B_0}{\hbar} & \f{iF}{\hbar}\\ \f{iF}{\hbar} & J-\f{B_0}{\hbar}\ea\right)\left(\ba{c}G_1\\G_2\ea\right)=0,
\eeq
where the operator $J$ is given by
\begin{eqnarray}
&&\ds J=-\f{d^2}{dx^2}+\f{B_0^2-F^2}{\hbar^2}(x+x_0)^2+\f{\Delta_0^2}{\hbar^2}\nonumber\\ && \phantom{xxxxxxxxxxxxxxxxxxxx}-\f{({\tilde\eps}_0B_0+\hbar kF)^2}{\hbar^2(B_0^2-F^2)},\nonumber\\
&& x_0=\f{{\tilde\eps}_0F+B_0\hbar k}{(B_0^2-F^2)}.
\end{eqnarray}
Since the operator $J$ is of the form of a displaced harmonic oscillator, the matrix differential equation (\ref{chi}) can be easily solved. Taking $G_1$ to be of the form
\beq
\ds G_{1n}\sim H_n\left(\sqrt{\f{\sqrt{B_0^2-F^2}}{\hbar}}(x+x_0)\right)~e^{-\f{\sqrt{B_0^2-F^2}}{2\hbar}(x+x_0)^2},
\eeq
the eigenvalues can be found to be
\begin{eqnarray}
\label{spectrum1}
&&\epsilon_n=\displaystyle-\tilde{E}\tilde{\theta}\hbar k v_F-\f{\tilde{E}\hbar k v_F(1-2\tilde{B}\tilde{\theta}+\tilde{\eta}\tilde{\theta})}{2(\tilde{B}-\tilde{\eta})}\nonumber\\
&&\phantom{\epsilon_n=}\displaystyle\pm \f{v_F}{2(\tilde{B}-\tilde{\eta})}\sqrt{\Delta_0^2\lam+\f{\hbar \lam^{\f{3}{2}}}{(1-\tilde{B}\tilde{\theta})^2}\,(2n)},\nonumber \\ &&\phantom{xxxxxxxxxxxxxxxxxxxx}n=0,1,2,\cdots,
\end{eqnarray}
where
\beq
\lam=4(1-\tilde{B}\tilde{\theta})^2(\tilde{B}-\tilde{\eta})^2-\tilde{E}^2(1-2\tilde{B}\tilde{\theta}+\tilde{\eta}\tilde{\theta})^2.
\eeq
The expression for the spectrum in (\ref{spectrum1}) is however not suitable for further analysis. Introducing the parameter $\displaystyle\beta={\tilde E}\g/B_0$ one may rewrite the above expression in a more convenient form as     
\begin{eqnarray}
\label{spectrum2}
\eps_n=&&\!\!\!\!\!\!\!\!\!-{\tilde E}{\tilde \theta}\hbar kv_F-\hbar kv_F\beta(1-\tilde B\tilde\theta)\phantom{xxxxx} \\ 
&&\!\!\!\!\!\!\!\!\!\pm \revision{v_F}(1-\tilde B\tilde\theta)\sqrt{\Delta_0^2(1-\beta^2)+\hbar B_0(1-\beta^2)^{3/2}(2n)}\nonumber.
\end{eqnarray}

\section{Analysis of the spectrum}
 \begin{figure}[t!]
\includegraphics[width=0.47\textwidth]{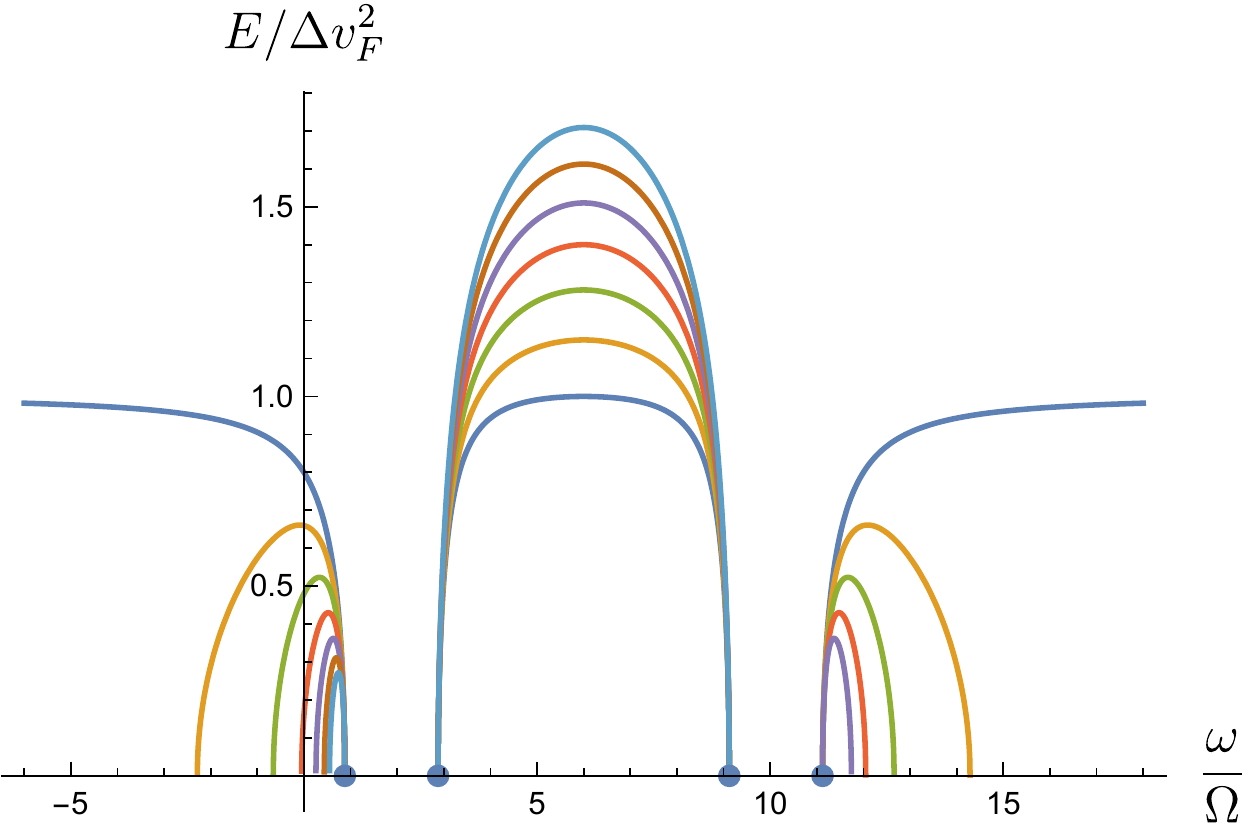} 
\caption{(color Online)\label{fig:spectrum} We show here the discrete part of the positive branch of the spectrum ($k=0$). We have fixed $\omega_\eta/\Omega=2$, and 
$\omega_\theta/\Omega=8$. The full dots  are the critical points $\omega_i$, ($i=1,2,3,4$ where $\beta \to 1$), as given in Eq.~(\ref{criticalpoints}a-d). It is clearly shown that at these points there is a collapse of the Landau levels. The different solid lines are for values of the quantum number $N=0,1,2,3,4,5,6$. 
}
\end{figure}
In this section we discuss and analyze the spectrum as given in Eq.~\eqref{spectrum2}.
With the definitions:
\begin{equation}
\label{definitions}
\begin{array}{ll}
\omega= \displaystyle\frac{\tilde{B}}{\Delta} = \displaystyle\frac{eB}{2\Delta c}, & \Omega =\displaystyle\frac{\tilde{E}}{2\Delta}=\displaystyle\frac{eE}{2\Delta v_F},  \\ \omega_\eta= \displaystyle\frac{\tilde{\eta}}{\Delta}=\displaystyle\frac{\eta}{2\hbar \Delta},& \omega_\theta= \displaystyle\frac{1}{\Delta \tilde\theta} = \displaystyle\frac{2\hbar}{\Delta \theta},
\end{array}\end{equation}
the factor $\beta$ becomes:
\begin{equation}
\beta=\Omega\frac{\omega_\theta -2 \omega +\omega_\eta}{(\omega-\omega_\eta)(\omega_\theta-\omega)} 
\end{equation}
and the roots, in the $\omega$ variable, of the equation $\beta=\pm 1$ are:
\begin{subequations}
\label{criticalpoints}
\begin{align}
\omega_1&= \frac{1}{2}\left(-2\Omega+\omega_\eta +\omega_\theta -\sqrt{4\Omega^2+(\omega_\eta-\omega_\theta)^2}\right),\\
\omega_2&= \frac{1}{2}\left(+2\Omega+\omega_\eta +\omega_\theta -\sqrt{4\Omega^2+(\omega_\eta-\omega_\theta)^2}\right),\\
\omega_3&= \frac{1}{2}\left(-2\Omega+\omega_\eta +\omega_\theta +\sqrt{4\Omega^2+(\omega_\eta-\omega_\theta)^2}\right),\\
\omega_4&= \frac{1}{2}\left(+2\Omega+\omega_\eta +\omega_\theta +\sqrt{4\Omega^2+(\omega_\eta-\omega_\theta)^2}\right),
\end{align}
\end{subequations}
where $\omega_1$ and $\omega_3$ are solutions of $\beta=-1$ while $\omega_2$ and $\omega_4$ are solutions of $\beta=+1$. Interestingly we note that the above critical values $\omega_i$ satisfy the following relations:
\begin{equation}
\omega_2-\omega_1 = 2 \Omega, \qquad \qquad \omega_4-\omega_3=2 \Omega. 
\end{equation} 
The above relations imply that even in the commutative limit $\eta, \theta \to 0$ ($\omega_\eta\to0, \omega_\theta\to\infty$) we are left with two distinct critical points: $\omega_{1,2}$. This means that each pair of critical points, $\omega_{1,2}$ and $\omega_{2,3}$ will coalesce into a  single critical point only at zero electric field or $\Omega=0$, c.f. Eq.~\eqref{definitions}.  

We easily see that:
\begin{equation}
\frac{\hbar B_0}{\Delta_0^2}= \frac{(\omega-\omega_\eta)(\omega_\theta-\omega)}{\Omega' \omega_\theta}, \qquad \Omega' =\frac{\Delta v_F^2}{2\hbar}
 \end{equation}
and the the energy levels can be written as:
\begin{eqnarray}
\label{spectrum3}
\displaystyle\frac{\epsilon_N}{\Delta v_F^2}\!\!\!\! &=&\!\!\!\! -\displaystyle\frac{\hbar k}{\Delta v_F}\left[ 2 \displaystyle\frac{\Omega}{\omega_\theta}+\beta \left(1-\displaystyle\frac{\omega}{\omega_\theta}\right)\right]\phantom{xxxxxx}\nonumber \\ &&\!\!\!\!\pm \sqrt{(1-\beta^2) +\displaystyle\frac{(\omega-\omega_\eta)(\omega_\theta-\omega)}{\Omega' \omega_\theta} (1-\beta^2)^{3/2} N},\nonumber
\end{eqnarray}
where $N=2n$ with $n=0,1,2$, $\dots$ 

Fig.~\ref{fig:spectrum} shows the discrete part of the spectrum ($k=0$) as a function of the quantity $\omega/\Omega \propto B/E$. The first thing to notice is that there are regions, namely, $\omega_1<\omega<\omega_2$ and $\omega_3<\omega<\omega_4$ where the energy eigenvalues are not real and so there are no bound states. An important feature of the spectrum is that in general there are three regions where bound states can be found: i) $\omega < \omega_1$; ii) $\omega_2<\omega<\omega_3$ and iii)   $\omega_4<\omega$. 
Another interesting point worth of notice is that  while in region ii) the energy eigenvalue is an increasing function of the quantum number $N$, in region i) and iii) the energy eigenvalue decreases when the quantum number $N$ increases. 
From Fig.~\ref{fig:spectrum} is is also clear that in the presence of non-commutativity the system still shows the collapse of the spectrum at all of the various critical values of the magnetic field ($\omega_{1,...,4}$) as it is found in the commutative scenario. 

\begin{figure}[t!]
\includegraphics[width=0.47\textwidth]{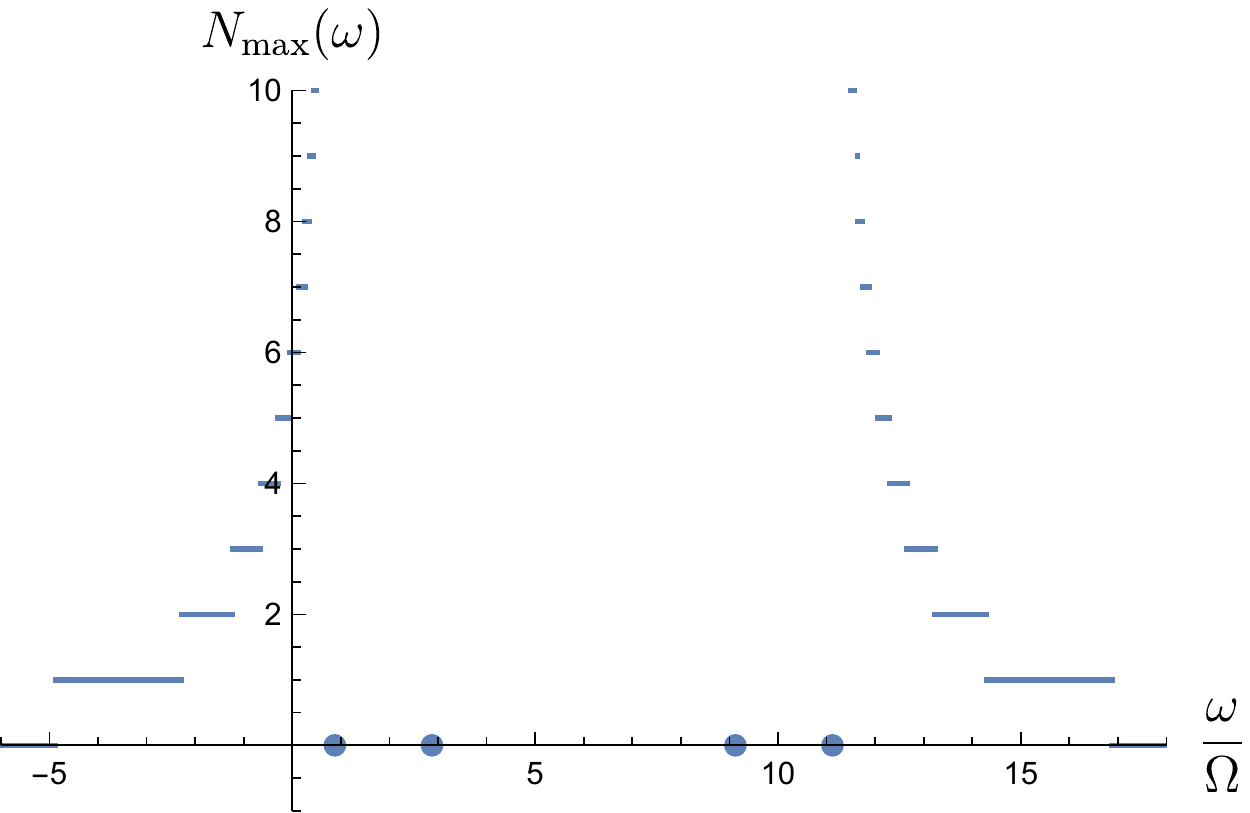} 
\caption{(color Online)\label{fig:FiniteNumber} We show here the quantity $N_{\text{max}}(\omega)$ as given in Eq.~\eqref{Nmax} in the different regions of the spectrum in $\omega$-space.  In the region (ii) $\omega_2 < \omega <\omega_3$ we have that $N_{\text{max}}=+\infty$ for any value of $\omega$.
}
\end{figure}

A notable feature is also that in region (i) and (iii), at each given value of  $\omega$ (that is of the magnetic field $B$) there is only a finite number of energy levels.  This is due to the fact that the factor multiplying the quantity $(1-\beta^2)^{3/2}$ in Eq.~\eqref{spectrum3} is negative in such regions while it is positive in region (iii).  Indeed there exist an $N_{\text{max}}(\omega)$ such that for any $N> N_{\text{max}}$ the discrete part of the energy eigenvalue becomes complex:
\begin{equation}
\label{Nmax}
N_{\text{max}}(\omega)= \text{IntegerPart}\left[\frac{\Omega' \omega_\theta}{(\omega-\omega_\eta)(\omega_\theta-\omega)\sqrt{1-\beta^2}}\right]. 
\end{equation} 
In addition as the critical points $\omega_1$ and $\omega_4$ are  approached respectively from below and from above such number of energy level increases indefinitely: $\lim_{\omega \to \omega_1^-}N_{\text{max}}(\omega) = \lim_{\omega \to \omega_4^+} N_{\text{max}}(\omega)=+\infty.$ 
Clearly this happens because $\beta\to 1 $ when $\omega$ approaches $\omega_1$ from below or $\omega_4$ from above. 
Fig.~\ref{fig:FiniteNumber} shows this behaviour. We can see  at the same time that as the critical points $\omega_1$  and $\omega_4$ are approached the  corresponding ``width'', in $\omega-$space, of the interval over which $N_{max}=\text{const}$  approaches zero.

On the contrary in the region $\omega_2 < \omega <\omega_3$ we have that $N_{\text{max}}=+\infty$. This can be seen easily by inspection of Eq.~\eqref{spectrum3}: the factor $(\omega-\omega_\eta)(\omega_\theta -\omega)  >0$ is in this region and thus $N$ is not bounded and can assume any integer value. 

It may be noted that the phase transition can also be described in terms of the electric field. We note for instance that the authors of ref.~\cite{Lukose} offer a treatment of the system in terms of a dual description in which at a fixed given value of the magnetic field a critical value of the electric field is found where the systems exhibits a collapse of the spectrum. We would like to remark here that in presence of non commutativity the situation is not symmetric  in the two approaches. Indeed we have seen in the above discussion that fixing the electric field one obtains four critical values of the magnetic field which are the roots of $\beta=\pm 1$ and this because $\beta$ is a quadratic function of the magnetic field ($\omega$).  On the contrary $\beta$  is linear in the electric field ($\Omega$) and at any fixed value of the magnetic field ($\omega$) there will only be two critical values  of the electric field, $E^{\text{crit}}_{\pm}$, (equal in magnitude and with opposite directions) with $\beta=\pm 1$:  
\begin{equation}
\label{Ecritgen}
E^{\text{crit}}_{\pm}= \left(\frac{2\Delta v_F}{e}\right)\Omega_{\pm}, \quad \qquad \Omega_{\pm}=\pm \frac{(\omega-\omega_\eta)(\omega_\theta-\omega)}{(\omega_\theta-2\omega+\omega_\eta)}\,.
\end{equation} 
One may notice that in the absence of non-commutativity, $\theta, \eta \to 0$ or $\omega_\theta \to \infty, \omega_\eta \to 0$, or $\Omega_\pm = \pm \omega$ and  the critical electric field is simply related to the magnetic field by :
\begin{equation}
E^{\text{crit}}_\pm= \pm \left(\frac{2\Delta v_F}{e}\right)\omega =\pm \frac{v_F}{c} B\, ,
\end{equation}
as discussed in ref.~\cite{Lukose}. 
When only momentum non commutativity is present ($\theta=0$, $\omega_\theta\to\infty$) we have:
\begin{equation}
\Omega_\pm= \omega -\omega_\eta
\end{equation}
and the critical electric field is shifted (reduced in magnitude) relative to the critical electric field of the commutative case.
When only coordinate non commutativity is present ($\eta=0$, $\omega_\eta=0$) we have:
\begin{equation}
\Omega_\pm= \frac{\omega}{1 -\frac{\omega}{\omega_\theta-\omega}}
\end{equation}
and the critical electric field is shifted (increased in magnitude) relative to the critical electric field of the commutative case (recall that we assume the natural ordering $\omega_\eta \ll \omega \ll \omega_\theta$). Of course in the general case of both non-commutative parameters non vanishing ($\theta,\eta \ne 0$) the net effect will be a combination of the two and given by Eq.~\eqref{Ecritgen}, and the reduction or increase of the critical electric field relative to the commutative scenario will depend on the relative strenght of the non-commutative parameters $\eta,\theta$ (or $\omega_\eta,\omega_\theta$).

\begin{figure}[t!]
\includegraphics[width=0.47\textwidth]{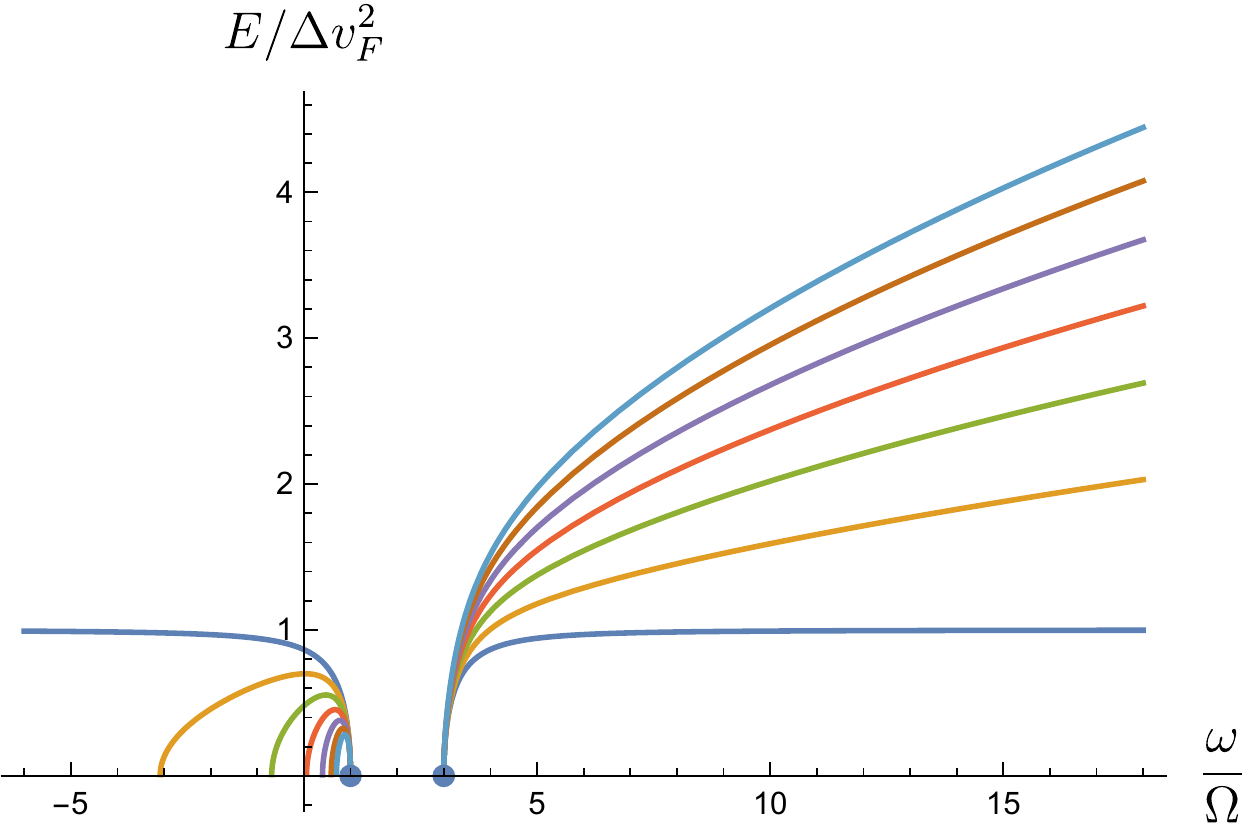} 
\caption{(color Online)\label{fig:spectrum2} We show here the discrete part of the positive branch of the spectrum ($k=0$). We have fixed $\omega_\eta/\Omega=2$, and 
$\omega_\theta/\Omega\to\infty$ ($\theta\to0$). The full dots  are the critical points $\omega_i$, ($i=1,2$ where $\beta \to 1$), as given in Eq.~(\ref{criticalpoints}a-b). It is clearly shown that at these points there is a collapse of the Landau levels. The different solid lines are for values of the quantum number $N=0,1,2,3,4,5,6$. The second critical region present in Fig~\ref{fig:spectrum} has moved off to $\omega\to \infty$.
}
\end{figure}
\begin{figure}[t!]
\includegraphics[width=0.47\textwidth]{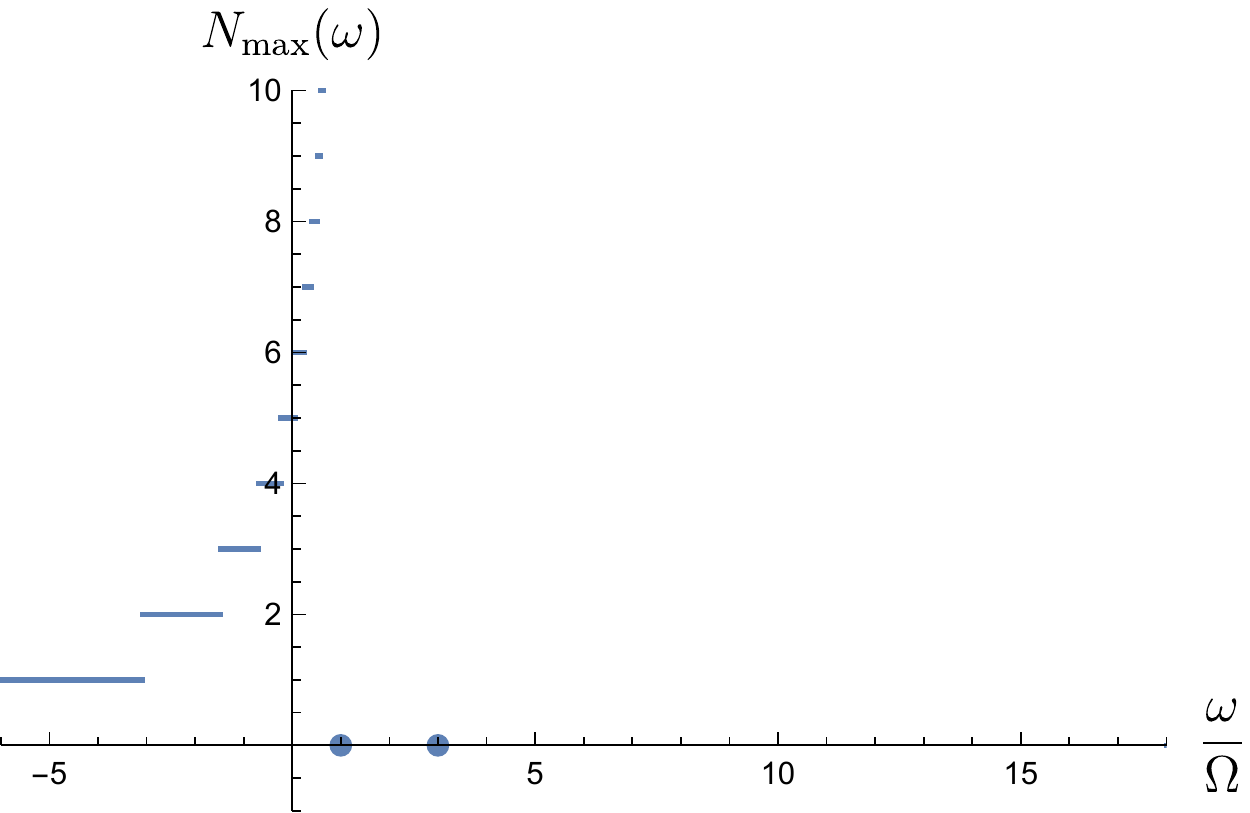} 
\caption{(color Online)\label{fig:FiniteNumber2} We show here the quantity $N_{\text{max}}(\omega)$ as given in Eq.~\eqref{Nmax} in the different regions of the spectrum in $\omega$-space when $\theta\to 0$.  In the region  $\omega > \omega_2 $ we have that $N_{\text{max}}=+\infty$ for any value of $\omega$.  The second critical region present in Fig~\ref{fig:FiniteNumber} has moved off to $\omega\to \infty$.
}
\end{figure}

Before  concluding we would like to make an important remark. While it is clear from the above discussion and from Fig.~\ref{fig:spectrum} and Fig.~\ref{fig:FiniteNumber} that the two critical regions, $\omega_1<\omega<\omega_2$ and $\omega_3<\omega<\omega_4$ share similarities an important feature is that they have a quite different behaviour when taking  the commutative limit. Indeed the second critical region  $\omega_3<\omega<\omega_4$ simply disappears when $\theta \to 0$, beacuse in this limit both $\omega_{3,4}\to \infty$, see Eqs.~\ref{criticalpoints}(c,d). On the contrary in the same limit $\theta \to 0$ the critical region $\omega_1<\omega<\omega_2$ will stand. Only if we also take the additional limit $\eta \to 0$ it will shrink to single  critical point $\omega_{1,2}=\mp\Omega+\omega_\eta/2$, see  Eqs.~\ref{criticalpoints}(a,b) which is then there also in the commutative limit. We also note that in the limit $\theta \to 0$ the net effect of momentum space non commutativity, $\eta \ne 0$, is simply to shift and split the single critical point $\omega=-\Omega$. Therefore we can say that the second critical region arises only when there is space coordinate non commutativity ($\theta \ne 0$).
For this reason we present the limiting form of the spectrum when $\theta\to 0$
\begin{eqnarray}
\!\!\!\!\!\!\!\!\!\!\!\!\left.\epsilon_n\right|_{\theta=0}\!\!\!&=&\!\!\!-\hbar kv_F\beta\pm \nonumber\\
&\phantom{x}& v_F\sqrt{\Delta_0^2(1-\beta^2)+\hbar B_0(1-\beta^2)^{3/2}(2n)}
\end{eqnarray} while in Fig.~\ref{fig:spectrum2} and Fig.~\ref{fig:FiniteNumber2} we present respectively the spectrum and the number of levels in the (two) critical regions in this limiting case of $\theta\to 0$ (i.e. only momentum non-commutativity). 

\section{Conclusions}

Although the effects of noncommutativity are generally associated with high energy physics, nevertheless such effects have been investigated within quantum mechanical models which are relatively easier to handle. As mentioned earlier the possible role of noncommutativity in condensed matter physics has also been studied by many authors. In view of this we have examined $(2+1)$ dimensional non-commutative Dirac equation in the presence of crossed magnetic and electric fields. It was shown \cite{Lukose} that in the commutative version of this system there is a critical electric field $E_{\text{crit}}$ such that for $E \to E_{\text{crit}}$ the Landau levels collapse to form a band. In this paper it has been shown that the same phenomena takes place in the non-commutative version and, as expected, the critical electric field depends on the parameters of non-commutativity. 

Apart from this another interesting phenomena takes place. It has been shown that if the electric field is kept fixed, there are multiple critical magnetic fields $B_{\text{crit}}$ such that for $B\to B_{\text{crit}}$ the spectrum shows contraction of Landau levels. It may be mentioned that these critical fields coalesce in the limit of $\theta,\eta\ra 0$. 
We have presented explicitly our results for the limiting case of only momentum non commutativity (i.e. coordinate space commutativity turned off, $\theta=0$, showing that in this limit the second critical region moves off to $\omega\to \infty$ while only the first critical region survives.


\acknowledgments 
One of us (P.~R.) wishes to thank INFN Sezione di Perugia for support. He would also like to thank the Physics Department of the University of Perugia for hospitality.

\end{document}